\def\K{{\rm K}}

\def\km{{\rm km}}
\def\s{{\rm s}}

\def\beq#1{\begin{equation}\label{#1}}
\def\eeq{\end{equation}}
\def\beqa#1{\begin{eqnarray}\label{#1}}
\def\eeqa{\end{eqnarray}}

\def\spose#1{\hbox to 0pt{#1\hss}}
\def\simlt{\mathrel{\spose{\lower 3pt\hbox{$\mathchar"218$}}
     \raise 2.0pt\hbox{$\mathchar"13C$}}}
\def\simgt{\mathrel{\spose{\lower 3pt\hbox{$\mathchar"218$}}
     \raise 2.0pt\hbox{$\mathchar"13E$}}}
\def\simpropto{\mathrel{\spose{\lower 3pt\hbox{$\mathchar"218$}}
     \raise 2.0pt\hbox{$\propto$}}}

\def\ed{\end{document}}

\def\P{{\rm P}}

\def\w{{\bf w}}

\documentstyle[emulateapj,danonecolfloat]{article}
\def\NoApjSectionMarkInTitle#1{#1.\ }
\begin{document}
\twocolumn[


\journalid{337}{15 January 1989}
\articleid{11}{14}

\submitted{\today. To be submitted to ApJ.}

\title{Searching for fluctuations in the IGM temperature using the
Lyman $\alpha$ forest}
\author{Matias Zaldarriaga}
\affil{Physics Department, New York University,
4 Washington Place, New York, NY 10003}
\affil{ 
Institute for Advanced Study, School of Natural Sciences,
Olden Lane, Princeton, NJ 08540}


\vskip 1pc


\begin{abstract}

We propose a statistical method to search for fluctuations in the
temperature of the intergalactic medium (IGM) using the Lyman $\alpha$
forest. The power on small scales ($\sim 25\ \km/\s$) is used as a
thermometer and fluctuations of this power are constrained. The method
is illustrated using Q1422+231. We see no evidence of temperature
fluctuations.  We show that in a model with two temperatures that
occupy comparable fractions of the spectra, the ratio of small scale
powers is constrained to be smaller than $3.5$ (corresponding to a
factor of $2.5$ in temperature). We show that approximately ten
quasars are needed constrain factors of two fluctuations in small
scale power power.

\end{abstract}


]


\section{Introduction}\label{introduction}

The Lyman $\alpha$ forest has become one of the major tools of
cosmology. It has been used to set constraints on many of the
parameters of the cosmological model, such as the amplitude and slope
of the spectrum of primordial density fluctuations and the nature of
the dark matter
(\cite{croft98,croft99,wc2000,naray00,zalhuiteg00,croft00}). In the
popular cosmologies, the forest arises rather naturally and both
cosmological simulation
(\cite{Cen94,her95,zhang95,m96,mucket96,wb96,th98}) as well as
analytical models have been used to understand its properties
(\cite{Bi92,rm95,BD97,GH96,croft97,HuiGne97,Hui97b}).

The statistical properties of the forest flux are affected by the
physical properties of the intergalactic medium (IGM) such as its
temperature or the slope of its equation of state affect. In turn
these physical properties of the IGM are sensitive to the ionization
history of the universe and the cosmological parameters. An intensive
effort has been devoted to understanding what different scenarios
predict for the forest and in constraining those scenarios with the
available observations
(eg. \cite{HuiGne97,gh98,joop99,McD00,croft00}).

The temperature of the IGM has been determined in two different ways,
by measuring the widths of the lines (\cite{joop99,McD00}) and by
studying the small scale power spectrum of the flux
(\cite{zalhuiteg00}).  Both methods find values of the temperature
around $T\sim 20,000 K$ which are higher than theoretical expectations
(ie. \cite{HuiGne97,Hui97b}). Both these type of studies as well as
most theoretical models assume that the temperature of the IGM is
constant in space. The aim of this paper is to formulate a method for
testing this assumption, illustrate it using the spectrum of one
quasar and investigate what type of constraints on the temperature
fluctuations could be obtained with more data.

A possible explanation for the high temperature of the IGM around
redshift $z \sim 3$ is heat deposited in the gas by the double
ionization of He. Depending upon the nature of the source responsible
for the reionization of the universe and its clumpiness, HeII
reionization can be substantially delayed relative to H reionization
and could happen around $z\sim 3$
(\cite{MiraldeRees,MadauMeiskin,EscudeHaenRees}).

A natural consequence of this scenario are fluctuations in the
temperature of the IGM around $z\sim 3$. In fact studies of the HeII
Lyman $\alpha$ forest show that the mean absorption is increasing
rapidly with redshift for redshifts between 2 and 3. Moreover several
gaps of size $\sim 1000 \km / \s$ where the transmitted flux is high
have been observed (\cite{Reimers97,Anderson98,Heap00,Smette00}).

A uniform IGM temperature has been assumed in most of the theoretical
and observational Lyman $\alpha$ forest work so far.  We live in an
inhomogeneous universe, inhomogeneous heating of the IGM due to a late
HeII reionization is only one of the possible sources of temperature
fluctuations. It is timely to investigate possible ways of detecting
temperature fluctuations. In this paper we use the small scale power
spectrum of the forest flux as a measure of temperature
(\cite{th00,zalhuiteg00}) an introduce the tools necessary to search for
fluctuations in the small scale power.

The paper is organized as follows, in section \S \ref{PS} we summarize
the properties of the small scale power spectrum and its dependence on
the IGM temperature, in \S \ref{wavelets} we introduce the tools we
use in the search for temperature fluctuations, in \S \ref{const} we
apply our technique to Q1422+231 (\cite{kim}), a quasar at redshift
$z=3.62$ and in \S \ref{future} we estimate how much data is needed to
substantially improve our constraints. We conclude in \S
\ref{conclusions}

\section{The power spectrum as a measure of the IGM temperature}\label{PS}

\begin{figure}[tb]
\centerline{\epsfxsize=9cm\epsffile{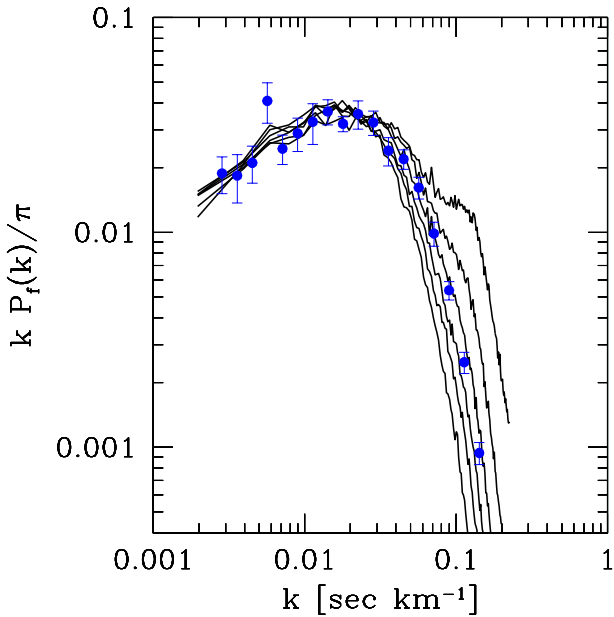}}
\caption{Transmission power spectrum from \cite{zalhuiteg00}. Models 
with have varying $T_0$ ($T_0= (150,250,\cdots 650) (\km/\s)^2$
increases from top to bottom, $\gamma$ was set to $0.2$ and the mean
transmission was set to $0.7$). The data points are from McDonald et
al. (1999).}
\label{movieps}
\end{figure}

The power spectrum of the flux in the Lyman $\alpha$ forest can be
used as an indication of the IGM temperature
(\cite{th00,zalhuiteg00}). Figure \ref{movieps} shows the
predictions of a set of models with different temperatures together
with the measurements of \cite{McD99}. The simulations used in this
paper are described in \cite{zalhuiteg00}. We use PM simulations to
solve for the dark matter and then use analytic scaling relations to
construct mock flux spectra. Our simulations have $128^3$ particles in
a box that at $z=3$ is $3200 \km/\s$ in size.

Given that the small scale power is sensitive to the temperature of
the IGM it is reasonable to try to detect fluctuations in the IGM
temperature by searching for fluctuations in the small scale power.
To do so we have to measure the power in different positions along the
spectrum and/or on lines of sight towards different quasars. We could
use Fourier analysis on pieces of the spectra but we prefer to use
wavelets, a natural tool to obtain measures of power that are
simultaneously localized in real and frequency space. In the next
section we review the properties of the wavelet expansion that are
relevant for our work.
 
\section{Wavelets}\label{wavelets}

Wavelets have been used in a number of statistical studies of the
forest. In \cite{meiskin00} wavelets were introduced for
data compression and several statistical properties of the wavelet
coefficients were presented.

For our purposes the more relevant work is that of
\cite{theunszaroubi00}. Just as we will do in this paper, the authors
used wavelet coefficients to measure the small scale power in the
spectrum. They presented the cumulative probability distribution of the squares
of the wavelet coefficients and focus most of their effort on
comparing models with different temperatures and equations of state
and thus different small scale power.  They conclude that the wavelet
coefficients in these different models have different enough
distributions to be able to tell the models apart observationally.  In
\cite{zalhuiteg00} we used Fourier analysis and measurements of the
small scale power in the literature to constrain the equation of state
of the IGM. Both \cite{theunszaroubi00} and \cite{zalhuiteg00} are
very similar in this respect, they propose the same ``thermometer''
for the IGM, the small scale power in the forest.
 
Furthermore, \cite{theunszaroubi00} argue that because wavelet
coefficients are localized in real space they can be used to measure
fluctuations in the temperature.  They showed results for the
cumulative distribution of a model with two temperatures and compare
them to models with one single temperature. Unfortunately it is
unclear how much of the difference is due to the models having a
different mean temperatures and how much is due to the fluctuations
themselves, or how much data is needed to get tight constraints on
fluctuations. Our work in this paper builds on the ideas presented in
\cite{theunszaroubi00} and \cite{zalhuiteg00} with special focus on
statistical tests of temperature fluctuations.
  
In this section we introduce the concepts necessary to understand
the use of wavelets in the context of this paper. 

\subsection{Definitions}

\begin{figure}[tb]
\centerline{\epsfxsize=9cm\epsffile{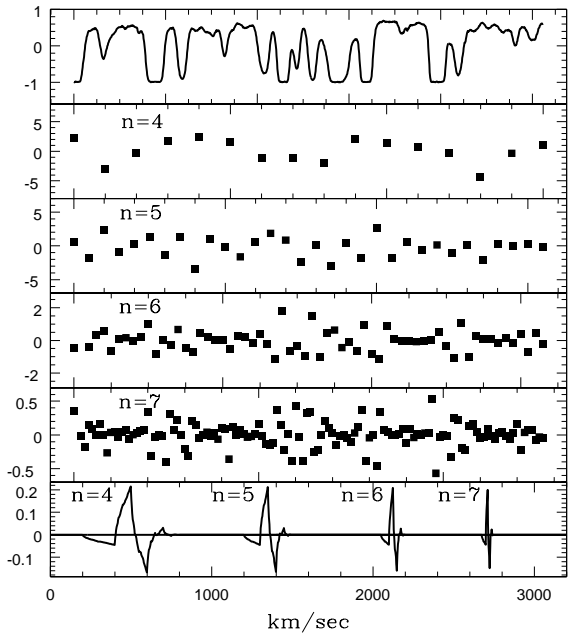}}
\caption{Wavelet transform of $3200 \km/\s$ of the spectrum of
Q1422+231. On the top we show the actual spectrum. In the middle
panels we plot the amplitude of the wavelet coefficients along the
spectra for four different orders. In the bottom panel we show
examples of the wavelet functions $\psi$ for the orders whose
coefficients are shown.}
\label{wavelets1}
\end{figure}

The best way to introduce the necessary concepts is by looking at the
wavelet decomposition of a piece of spectra from Q1422+231
(\cite{kim}). In figure \ref{wavelets1} we show $3200\ \km/\s$ of the
spectra together with wavelet coefficients. This is approximately a
10th of the Lyman $\alpha$ part of the spectra. We will use all the
spectra later, here we chose to show a small portion mainly for
clarity but also because this is the length of the simulations we will
have to compare with.  The wavelet transform of the spectra is
obtained by convolving the spectra with a set of test functions which
we will call $\psi_n(x-x_i)$, 
\beq{ampdef} a_n(x_i)=\int dx F(x)
\psi_n(x-x_i), 
\eeq $x$ 
denotes position along the spectra and $F(x)$
is the flux.  The test functions are characterized by two numbers, the
index $n$ and $x_i$. Increasing values of $n$ correspond to higher
spatial frequencies, $x_i$ determines the position of the wavelet
along the spectrum. The values of $x_i$ are determined by the wavelet
scheme. For the wavelets we are using there are $2^n$ values of $x_i$
in each order $n$.  Figure \ref{wavelets1} shows a few examples of
these wavelets to illustrate how as $n$ increases, the wavelet
corresponds to higher spatial frequencies. In this work we use the
Daubenchies 4 wavelets implemented in a Numerical Recipes routine
(\cite{press}).

For our purposes the wavelets are just a set of filters that we apply
to the data. They have the advantage of being localized in both
Fourier and spatial domain,  of being an orthogonal and complete set,
and perhaps most importantly that packages that perform the transform
are readily available.  

\begin{figure}[tb]
\centerline{\epsfxsize=9cm\epsffile{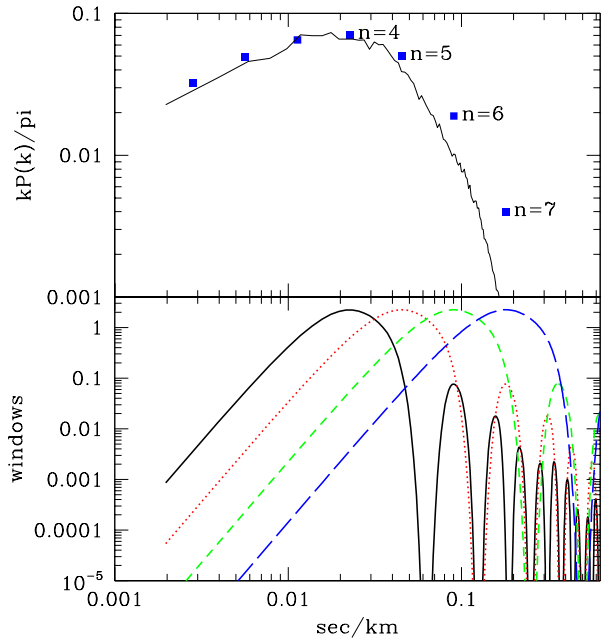}}
\caption{In the top panel we show the power spectrum of the forest in
a cosmological simulation measured both using Fourier and wavelet
analysis. The line corresponds to Fourier analysis, the points are the
equivalent measures using wavelets. On the bottom we show the window
functions for the $n=4,5,6,7$ wavelets. The power measured using
wavelets is a weighted average of the Fourier power spectrum with the
windows as weights.}
\label{powersp}
\end{figure}

To further de-mystify the wavelets we can look at the power spectrum of
the forest from a wavelet perspective. Rather than computing squares
of Fourier coefficient we can obtain a measure of power by averaging
the squares of the wavelets coefficients of a fixed $n$ but different
positions along the spectrum. We have
\beq{powerwav}
\hat P \propto {1\over N} \sum_{i=1}^N a_n^2(x_i),
\eeq
the constant of proportionality depends on the normalization convention
of $\psi_n$. The wavelets are well localized in real space so they
must have a broad response in Fourier space. These response 
or window functions for a few orders are shown in figure
\ref{powersp}. By construction window functions of successive $n$
differ only by a factor of 2 scaling in the $x$ axis. In figure
\ref{powersp} we also show the power spectrum measured in the
conventional Fourier way or using wavelts. The $x$ coordinate of each
wavelet measurement is set to the peak of the window function, but the windows are
clearly very broad.

We should point out that the $n$ label is arbitrary in the sense that
the label for a wavelet that probes a particular range of spatial
scales depends on the length of the spectrum being analyzed. For
example, if we start with a piece of spectrum that is twice as long
($6400\ \km/\s$) all the labels would be shifted by one. That is, the
new $n=7$ wavelet probes the same scales as the old $n=6$ did. As long
as one is consistent with the normalization convention the new wavelet
coefficients of order $n+1$ will be equal to the old coefficients of
order $n$ at the same position along the spectrum. To avoid confusion
we will use the labels corresponding to $3200 \km/\s$ in all our
discussions even when we analyze the full length of Q1422+231
spectrum. Figure \ref{powersp} indicates exactly what scales we probe
when we discuss each order $n$.

\begin{figure}[tb]
\centerline{\epsfxsize=9cm\epsffile{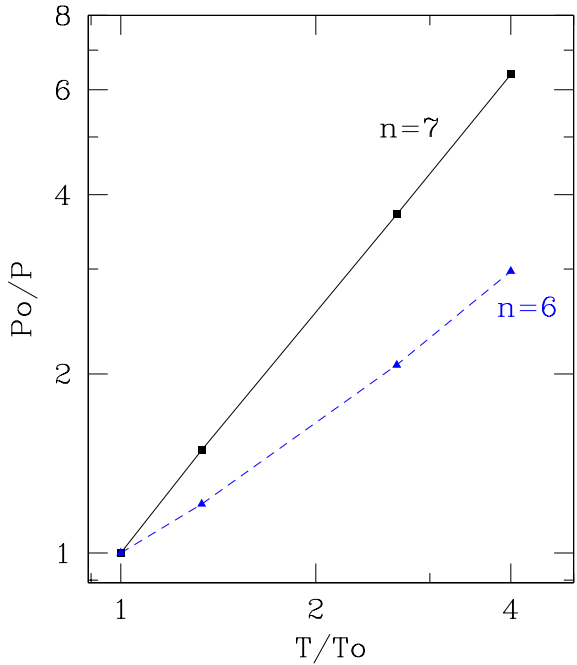}}
\caption{Ratio of power (average of the square of the wavelet
coefficients) for models with different temperatures normalized to a
model with $T=300 (km/\s)^2=1.8\times 10^4 \K$ in our simulations. We
show the results for $n=6$ and $n=7$ wavelets.}
\label{powervstemp}
\end{figure}

It is useful to discuss how the power in the wavelet coefficients
depends on the temperature. In figure \ref{powervstemp} we illustrate
the rate of change of power with temperature obtained in our
simulations. Clearly the smallest spatial scales are most
sensitive. The figure also shows that for $n=7$ a factor of $2$ change
in the temperature, which could be expected in models were He has been
doubly ionized in patches of the spectrum, could lead to a factor of
$2.5$ change in the observed power. For the figure me assumed an
isothermal model, the temperature of the gas is independent of its
density.  In \cite{zalhuiteg00} we showed that the small scale power
is sensitive to the temperature over a range of overdensities $ 0.9 <
\rho/\bar \rho < 1.8$. The ratio of temperatures in figure
\ref{powervstemp} should be interpreted as the ratio of temperatures
on that range of overdensities. This fact can be particularly
important in the case of energy injection by He reionization as the
process will not only change the temperature of the gas but also its
equation of state, so the relevant parameter is by how much the
temperature is increased in the range of overdensities that are probed
by the small scale power.

In summary wavelets of different scales $n$ measure power on different
spatial frequencies. Their frequency response is broad, the trade-off
for the wavelets being localized in real space. The wavelets
probing the small scales are very sensitive to the temperature of the
IGM and we will use these coefficients as probes of the temperature.
We will take advantage of the localization in real space of the
wavelets to search for temperature fluctuations.

\subsection{Statistical properties of the wavelet coefficients}
 
Now that we have defined the wavelet coefficients we will turn to some
analysis of their statistical properties in the case of the Lyman
$\alpha$ forest flux.

\begin{figure}[tb]
\centerline{\epsfxsize=9cm\epsffile{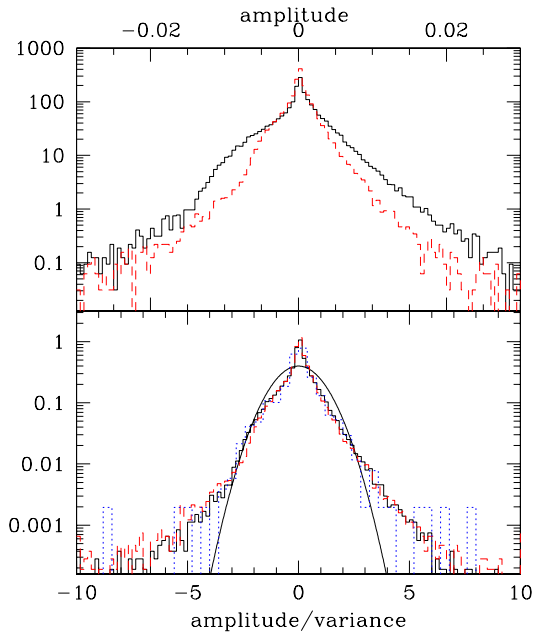}}
\caption{In the top panel we show the histograms of the amplitude of $n=7$
wavelet coefficients in our $3200 \km/\s$ simulations for two
different temperature,  $T=400 (\km/\s)^2$ (larger variance) and $ T=800 (\km/\s)^2$
(smaller variance). In the bottom panel we have rescaled the axis
using the variance. We also show the distribution for the coefficients
of $Q1422+231$ (dotted line) and a simple Gaussian (solid line).}
\label{histo}
\end{figure}

In the rest of the paper we will be interested in the small scale
power, that we will be using as a thermometer. In figure \ref{histo}
we show a histogram of the amplitude of the $n=7$ coefficient
obtained in our simulations. The variance of this distribution
proportional to the flux power spectrum (equation \ref{powerwav}). In
the the top panel we show the histograms for two different
temperatures, $T=400\ (800)\ (\km/\s)^2 = 2.4\ (4.8)\times 10^4
\K$. Clearly the model with the largest temperature has the smaller small
scale power, the variance of the distribution is smaller.

In the bottom panel of figure \ref{histo} we have rescaled the $x$
axis by dividing it by the variance. The two histograms now fall on
top of each other, although there are some small differences. For
comparison we also show the histograms of the coefficients of
Q1422+231 which have a distribution which is remarkably similar to the
coefficients in the simulation.

In figure \ref{histo} we show a Gaussian together with the histograms
of the coefficients in the simulations and in Q1422+231. There are
clear differences between the Gaussian and the rest. The amplitudes of
the wavelets of the flux have a probability of being near zero which
is much larger than one would expect in a Gaussian of the same
variance. This is clearly noticeable in the example of figure
\ref{wavelets1}, several stretches of the spectrum can be identified
where the amplitude of the wavelet coefficients are very small. The
distribution of the coefficient on the simulation also has tails,
it is much probable to get large coefficients than in a Gaussian
distribution.

\begin{figure}[tb]
\centerline{\epsfxsize=9cm\epsffile{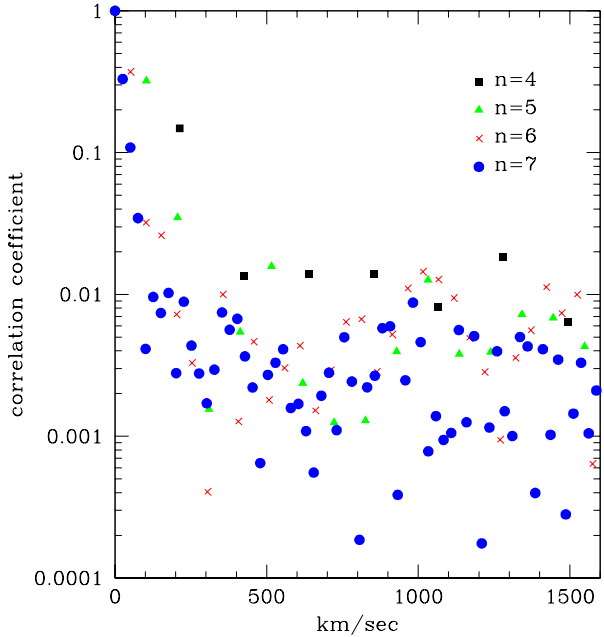}}
\caption{Correlation coefficient for wavelets of different indices $n$
as a function of separation $\delta v$. For the smallest scales ($n
\leq 6$ the correlations become negligible ($< 1\%$) foe separations
larger that $\sim 100 \km/\s$.}
\label{corrcoeff}
\end{figure}

To lay the foundation for our method to probe for temperature
fluctuations we also need to study the spatial correlations of the
wavelet coefficients. In figure \ref{corrcoeff} we show the
correlation coefficient of the wavelet amplitudes ($CC$) as a function
of separation along the spectra ($\delta v$) for several scale indices $n$,
\beq{cc} 
CC(\delta v)={<a_n(x)a_n(x+\delta v)> \over <a_n^2>}.  
\eeq
The correlation coefficient decays fast with separation and then levels
off at a value below $1\%$. In our future analysis we will be
considering only the smaller scales, corresponding to $n=6$ or
$7$. The correlations in these cases become negligible (below
$1\%$) for separations larger than $\sim 100 \km/\s$.

\subsection{Analytic tools}

In order to detect temperature fluctuations we want to look for
variations in the statistical properties of the wavelet coefficients
as we look at different parts of the spectra or as we compare
different lines of sight. It is clear that any statements about
fluctuations will be statistical in nature, the variance in the
distribution of the wavelet coefficients is our thermometer, not the
coefficients themselves. Thus we cannot measure the temperature on a
pixel by pixel basis, we need to average over many pixels.

\begin{figure}[tb]
\centerline{\epsfxsize=9cm\epsffile{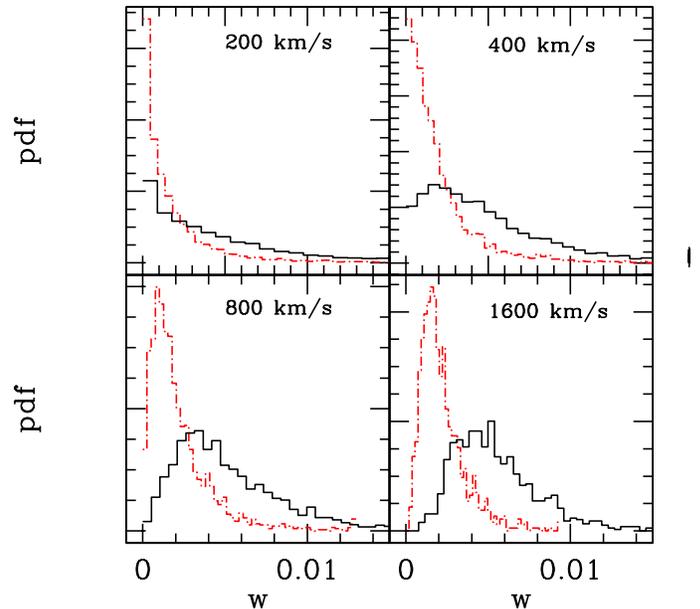}}
\caption{Probability distribution function of $w$ (defined in equation
\ref{wdef}) for four different choices of $L$ and $n=7$. Each panel
shows two temperatures,$T=400 (\km/\s)^2$ (larger mean) and $ T=800
(\km/\s)^2$ (smaller mean). The mean of $w$ is independent of $L$, it
depends on the temperature and $n$.}
\label{pdfw}
\end{figure}

Our measure of power is the square of the wavelet coefficient, thus we
will introduce averages of the square of these coefficients which we
will call $w$, 
\beq{wdef} w={1\over N_p}\sum_{i}a_n^2(x_i).  
\eeq 
The sum is done over contiguous portions of the spectra of length
$L$. Different portions do not overlap, and there are $N_p$ coefficient
in each chuck of size $L$. $N_p$ depends on the order $n$ one is
considering, $N_p= 2^n (L/3200 \km/\s)$. By definition the mean of $w$
is the measure of power and thus temperature. The mean only depends on
the index $n$ and not on $L$. In figure \ref{pdfw} we show the
histogram or probability distribution (PDF) of $w$ for several choices
of $L$ in our simulation. We show the histograms for two
temperatures. It is clear from the figure that the model with the
larger temperature has a smaller mean $w$.

\begin{figure}[tb]
\centerline{\epsfxsize=9cm\epsffile{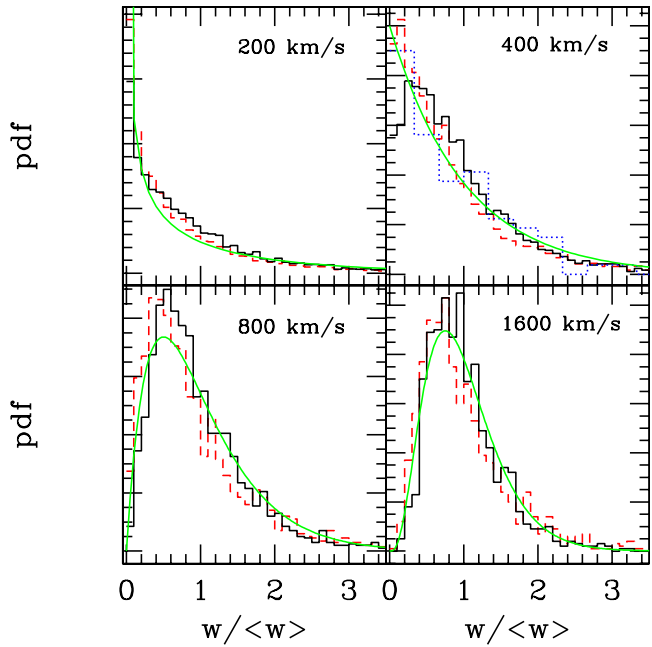}}
\caption{Rescaled probability distribution function of $w$ 
for four different choices of $L$ and $n=7$. Each panel shows two
temperatures, $T=400 (\km/\s)^2$ (solid) and $ T=800 (\km/\s)^2$
(dashed) together with our analytic model.  In addition in the $L=400
\km/\s$ panel (the smoothing scale we use in our statistical study) we
show the pdf for Q1422+231. }
\label{pdfwres}
\end{figure}

In figure \ref{pdfwres} we have rescaled the $x$ axis in figure
\ref{pdfw} using the mean of $w$ ($\bar w$) for each temperature.  The
histograms for the two temperatures are quite similar, we are thus
encouraged to find some simple analytic model.  Our statistic $w$ is
an average of squares of many variables which as figure
\ref{corrcoeff} shows quickly become uncorrelated. We could expect
that a $\chi^2$ distribution would be a good model for the observed
distributions. The effective number of degrees of freedom $\nu$ of the
$\chi^2$ distributions need to be determined. Furthermore because the
different panels correspond to values of $L$ that increase by factors
of 2, the effective number of degrees of freedom should increase by
factors of 2. In figure \ref{pdfwres} we plot this simple analytic
approximation,
\beqa{pdfanal} 
P(w|\bar w)&=&{\nu \over \Gamma(\nu/2) 2\bar w} \ \
e^{-u}\ \ u^{\nu/2-1} \nonumber \\ u&=& {w\nu/2 \bar w}, 
\eeqa 
where for the different panels we choose $\nu=1,2,4,8$. Changing
slightly the values of $\nu$ improves or worsens the agreements
depending on the panel. We choose these values doing a simple $\chi$
by eye of the above plots. Our constraints on the temperature
fluctuations are insensitive to the details of this choice, different
values of $\nu$ give the same answer, as long as the analytic formulas
fit the panels reasonably well. Note also that the distribution for
the data of Q1422+231 is also quite similar, there are some
differences for the largest smoothing window but in that case there
are only a few samples Q1422+231, so the discrepancy is not
significant.

We will use this analytic form for the pdf of $w$ throughout the rest
of our work. Although the approximation is not perfect our analytic
formulas give a good description what is observed in the
simulations. We are going to be searching for variations in this
distribution as we move along the spectra, so it is very useful that
we have an analytic form and furthermore that this analytic form
depends on only one parameter $(\bar w)$.  When formulated this way, the question
of fluctuations in the temperature can be very simply stated. Does a
model with a single value of $\bar w$ fit the data better than a model
were $\bar w$ varies in some particular way across the spectrum.

We could take a different approaches to look for fluctuations in the
statistical properties of the coefficient. We could directly compare
the observed distributions with what is obtained in simulations where
a uniform temperature was assumed and search for differences. The
problem with this approach is that one must establish that this
differences are not a consequence of systematic effects introduced in
the simulation process by some missing physics or some numerical
artifact. We believe that our method is more robust as it does not use
simulations directly, only to motivate our choice of distribution. We
are essentially looking for changes in the parameters of the
distribution we use to model the data as we go along the spectrum. Any
change in this parameters is a signal of fluctuations even if our
distribution is not a perfect model of the data.

\section{Constraints on temperature fluctuations}\label{const}

Our objective is to use the small scale power as a thermometer, and
this thermometer to search for fluctuations. In the
previous sections we have established a model  for the probability
distribution of the square of the wavelets coefficients. This model is
dependent on one parameter, which we chose to be the mean power on
that scale. The objective of this section is to describe the formalism
necessary to constrain spatial fluctuations in the mean power, such as
those that would be produced by fluctuation in the temperature. 

Let as call $\w$ the vector with the square of the amplitude of the
wavelet coefficient as a function of position along the spectra. Let
us denote $N$ the dimension of that vector, which is given by the
ratio of the length of the spectrum to length of the averaging filter
in equation (\ref{wdef}). For simplicity we will call each of the
entries of these vectors pixels, although they are a combination of
several pixels in the initial spectrum. 

In the previous sections we found a model for $\P (w_i| \bar w_i)$,
the probability of measuring $w_i$ given that mean $\bar w_i$ of $w$
at position $i$ along the spectrum, (equation \ref{pdfanal}). To
describe our model for the fluctuation we need to introduce another
vector ($\bar \w$) that specifies the value of $\bar w$ along the
spectrum. The dimension of $\bar \w$ is also N.  The probability of
measuring the vector $\w$ is given by,
\beqa{prob}
\P(\w)&=&\int d\bar \w \P(\bar \w) \P (\w| \bar \w) \nonumber \\ 
&=&\int d\bar w_1 \cdots d\bar w_n \P(\bar \w) 
\prod_{i=1}^N \P (w_i| \bar w_i). 
\eeqa
Different models for the temperature distribution differ in what
$\P(\bar \w)$ is. For example for a uniform temperature model where for
every pixel the mean power is $\bar w$ we have, 
\beq{unif1}
\P(\bar \w)=\delta^D(\bar w_1-\bar w)\cdots \delta^D(\bar w_N-\bar w)\
\eeq

By combining equations (\ref{pdfanal}), (\ref{prob}) and (\ref{unif1})
we obtain the probability of measuring $\w$ under the assumption
that the temperature is uniform ($\P^u$),
\beqa{punif}
\ln \P^u&=& \sum_i \ln \P(w_i|\bar w) \nonumber \\
&=& \sum_i (\nu/2-1) \ln(u_i)-u_i \nonumber \\
&-& N \ln [ 2 \bar w
\Gamma(\nu/2)/\nu], 
\eeqa
where $u_i=w_i\nu/2\bar w$.

We want to constrain fluctuations in the temperature by comparing the
probability of having measured a vector $\w$ in a particular model for
the temperature fluctuations ($\P^f$) with the same probability in
a uniform temperature model ($\P^u$). The ratio of $\P^f$ and $\P^u$
is the likelihood ratio for the two models. In order to compute $\P^f$
we need to specify the value of $\P(\bar \w)$, our model for the
temperature fluctuations. Clearly there are many different choices of
$\P(\bar \w)$ which describe different physical scenarios and it would
be impossible for us to try every possibility. Instead in this section
we have chosen some particular examples and found constraints in
those scenarios.

In all our models we will assume that there are only two different
temperature values possible and that the mean power for each of these
two temperatures $\bar w_A$ and $\bar w_B$, differ by a factor
$\beta=\bar w_B/\bar w_A$. We will also define $f$ as the fraction of
the spectrum with mean $\bar w_A$. The mean power averaged over all
pixels becomes,
\beq{avw}
\bar w = f \bar w_A + (1-f) \bar w_B.
\eeq
When comparing different models we will keep this mean, $\bar w$
fixed to the observed value in the spectrum. 
We are not interested in comparing models with different mean power
but models with or without spatial temperature fluctuations in the
power. 

There are two probability ratios that define the problem,
\beqa{pgwnu}
\ln {\P(w_i|\bar w_A)\over\P(w_i|\bar w)} &=& - u_i (1-f)(1-\beta)
\nonumber \\ &+& {\nu \over 2} \ln
(f+\beta (1-f)) \nonumber \\
\ln {\P(w_i|\bar w_B)\over\P(w_i|\bar w)} &=& - u_i f {(\beta -1)\over \beta}
\nonumber \\ &+& {\nu \over 2} \ln
({f+\beta (1-f) \over \beta}), 
\eeqa
where $u_i=w_i\nu/2\bar w$. The expressions for $\P^f/\P^u$ in the
different models for the fluctuations will always be expressed in
terms of these two quantities.

\subsection{Independent pixels}

We first consider a model in which every pixel is independent of the
rest and can be in one of two temperatures, corresponding to mean
powers $\bar w_A$ and $\bar w_B$. The prior on the small scale power becomes,
\beq{indep1} 
\P(\bar \w)=\prod_i [f \delta^D(\bar w_i- \bar w_A) + (1-f)
\delta^D(\bar w_i- \bar w_B)]. 
\eeq 
The likelihood ratio $\P^f/\P^u$ is, 
\beqa{indep2} 
\ln [\P^f/\P^u]
&=& \sum_i \ln [ f {P(w_i | \bar w_A) \over
P(w_i | \bar w)} \nonumber \\
&+& (1-f) {P(w_i | \bar w_B) \over P(w_i | \bar w)}]. 
\eeqa
This expression remains constant if we transform $\beta \rightarrow
1/\beta$ and $f\rightarrow (1-f)$.

\begin{figure}[tb]
\centerline{\epsfxsize=9cm\epsffile{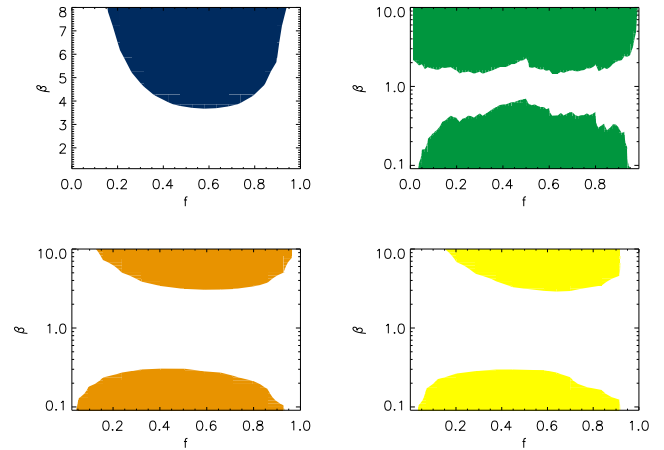}}
\caption{Two dimensional constraints on the fraction $f$ of spectrum with
mean $\bar w_A$ and ratio $\beta=\bar w_B/ \bar w_A$. The dark regions
have a small likelihood ratio, $\P^f/\P^u < 0.05$. The panels
correspond to different models for the fluctuation, independent pixels
(top left), sharp transition (top right), fixed size bubbles 
of size $1200 \km/\s$ (lower left) and of size $2400 \km/\s$ (lower right).}
\label{2dconst}
\end{figure}

We use equation (\ref{indep2}) and the full spectra of Q1422+231 to
put simultaneous constraints on $f$ and $\beta$ which are shown in
figure \ref{2dconst}. We do not need to consider values of $\beta<1$
because they can be obtained from $\beta >1$ using the symmetry
property of the likelihood ratio.  We considered models with a
likelihood ratio smaller than 0.05 as ruled out. As can be expected
the tightest constraints on $\beta$ are obtained when $f \sim
0.5$. With the data we have we can say that $\beta < 3.5$ ($f\sim 0.5$). 

As could have been guessed the tightest constraints are obtained when
both temperatures occupy a significant fraction of the spectrum,
$f\sim 0.5$.  However it is interesting to note, that the contour is
not symmetric around $f=0.5$. Figure \ref{2dconst} shows that it is
harder to put constraints when $f$ is near zero than when it is near
one. The reason for this is that when $f\sim 0$ and $\beta > 1$, most
of the spectrum has the larger small scale power $\bar w_B$ (low
temperature) and only a small fraction has the lower power $\bar w_A$
(high temperature). Even when $\bar w$ is high, figure \ref{pdfwres}
shows that it is more likely that in any given pixel the value of $w$
is low, thus it is hard to detect the small excess of pixels with low
$w$s that is produced by the small fraction of the spectrum that
has $\bar w_A$.

\subsection{Sharp transition}

We now consider a case in which there is a sharp transition
between $\bar w_A$ and $\bar w_B$ which occurs at pixel number $i_{tr}$. The
prior becomes,
\beq{sharp1} 
\P(\bar \w)=\prod_{i<i_{tr}} \delta^D(\bar w_i- \bar w_A)  \prod_{i>i_{tr}} \delta^D(\bar w_i- \bar w_B).  
\eeq 
The likelihood ratio $\P^f/\P^u$ for this case becomes, 
\beqa{sharp2} 
\ln [\P^f/\P^u]
&=& \sum_{i<i_{tr}} \ln [{P(w_i | \bar w_A) \over
P(w_i | \bar w)}] \nonumber \\
&+& \sum_{i>i_{tr}} \ln [{P(w_i | \bar w_B) \over P(w_i | \bar w)}]. 
\eeqa
For this model we define $f=i_{tr}/N$. We compute $\P^f/\P^u$ 
for all possible values of $i_{tr}$ and plot
the constraints on $\beta$ in figure \ref{2dconst}. We plot the
results in terms of $f$ rather than $i_{tr}$. For this model we 
plot constraints from both $\beta$ bigger and smaller than one, as
both cases are not equivalent. One corresponds to a jump from low 
to high temperature and the other to the opposite case.

These model is the one where the tightest constraints can be obtained
because it is the less random. Given $i_{tr}$ the temperature of all
pixels is specified. We can can constrain $0.5 \le \beta \le 2$ for almost all
values of $f$. 

\subsection{Fixed size bubbles}

Finally we will consider a case in which there are bubbles of a fixed
size with power $\bar w_B$ and the rest of the spectrum has $\bar
w_A$.  To compute the $\P (\w)$ under this scenario we reinterpret
equation (\ref{prob}) as an average over $\P (\bar \w)$ of $\P (\w|
\bar \w)$. We can use a Montecarlo technique to perform this
average. We create realizations simply by laying down $N_{b}$ bubbles
of a fixed size in the spectrum. We choose randomly the centers of the
bubbles. For each realization we compute the likelihood using equation
(\ref{pgwnu}) and the average that over all realizations to compute
the $\P (\w)$. The model has two free parameters, the size of each
individual bubble and the number of bubbles (or equivalently $f$, the
fraction of the spectrum having $\bar w_A$ and $\bar w_B$).

In figure \ref{2dconst} we show the constraints on $f$ and $\beta$ for
two particular bubble sizes, $1200 \ \km/s$ and $2400 \ \km/s$. In
this model $\beta$s larger and smaller than one are not
equivalent. Our results for independent pixels can be interpreted as a
constraint when the typical size of the bubble is $400 \ \km/s$. As
can be seen in figure \ref{2dconst} our constraints end up being very
similar for all bubble sizes. For $f\sim 0.5$ the ratio of powers have
to be in the range $0.3 \le \beta \le 3.5$.

\section{Future}\label{future}

In the previous sections we have developed a method for constraining
temperature fluctuations. We illustrated our method using data from
Q1422+231. The constraints on the ratio of the powers obtained are
somewhat larger than what one might expect to be present in the
universe. In this section we compute the number of spectra needed to
substantially improve the constraints.

The statistical question we will ask is what constraints on $\beta$
can be obtained from a certain amount of data in the independent pixel
model under the assumption that the universe has uniform temperature.
Equation (\ref{indep2}) implies that the ratio of the probabilities of a
uniform model to the fluctuating temperature model ($\P^{f}/\P^{u}$)
is,
\beqa{future1} 
\ln (\P^{f}/\P^{u}) &=& \sum_i \ln [ f {P(w_i | \bar w_A) \over
P(w_i | \bar w)} \nonumber \\ 
&+& (1-f) {P(w_i | \bar w_B) \over P(w_i | \bar w)}] \nonumber \\
&\approx& N \int dw \P(w|\bar w) \nonumber \\
&\ln& [ f {P(w | \bar w_A) \over
P(w | \bar w)} 
+(1-f) {P(w | \bar w_B) \over P(w | \bar w)}]
\eeqa
In the last line we approximated the sum by a mean over the distribution
from which the $w_i$ are assumed to be drawn, $\P(w|\bar w)$, and we
used their independence. The last line of equation (\ref{future1}) is
a measure of distance between two probability distribution functions
called Kullback-Leibler entropy (\cite{kull59}).

\begin{figure}[tb]
\centerline{\epsfxsize=9cm\epsffile{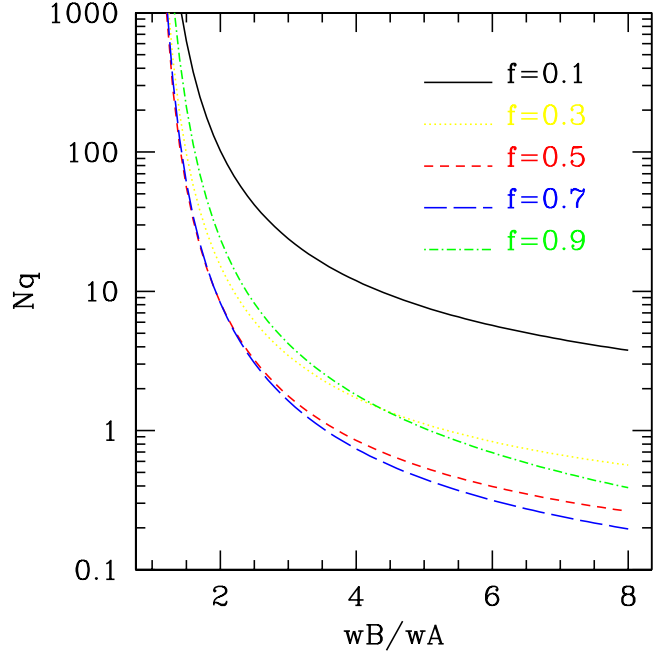}}
\caption{Number of quasars similar to Q1422+231 needed to obtain a
constraint for several values of $f$ as a function of $\beta=\bar
w_B/\bar w_A$.}
\label{futurefig}
\end{figure}

We can set the ratio in equation (\ref{future1}) to $0.05$ and solve
for $N$ as a function of $\beta$ and $f$.  We show the results in
figure \ref{futurefig}. We plot only points for $\beta > 1$, other values
of $\beta$ can be obtained using the transformation $\beta \rightarrow
1/\beta$, $f\rightarrow (1-f)$.  We have changed variables form $N$ to
$N_q$, the number of quasars with the same number of pixels as
Q1422+231.  The figure shows that ten quasars are needed to obtain
constraints on $\beta\sim 2$ for $f\sim 0.5$. The figure also
illustrates the fact that it is very difficult to obtain constraints
that are much tighter than a factor 2 in $\beta$.

It is interesting to understand why it gets so much harder to obtain
constraints when $\beta$ approches one. Figure \ref{futurefig} shows
that the number of quasars scales as $N_q \propto 1/(\beta-1)^4$. The
key point is that we are comparing models that by construction have
the same mean $\bar w$ (the same mean temperature). Thus what we are
actually comparing is the variance of the $w$ distribution in both
models. We can compute how many samples are needed to distinguish two
models that have different variance. We define the variance as
$v=<w^2>-\bar w^2$, its value depends on the model (fluctuating or
uniform temperature), but $\bar w$ is the same across models. We are
assuming all the pixels are independant so we can estimate the
variance as,
\beq{vardef}
\hat v={1 \over N_p}\sum_i w_i^2- \bar w^2.
\eeq

The question becomes how many samples $(N_p)$ are needed to distinguish a
model with uniform temperature with one characterized by $\beta$ and
$f$.  We define the signal to noise ratio as,
\beq{s2n}
(S/N)^2 ={[v^f-v^u]^2/{\rm var}(v)},
\eeq
the ratio of the difference in variance between the model with
fluctuations and the one without to the variance of the variance
(which we calculate in the uniform model).

In our example the signal to noise ratio becomes,

\beq{scaling}
(S/N)^2 =c N_p  {(1-\beta)^4 f^2 (1-f)^2 \over [f+\beta (1-f)]^4},
\eeq
where the constant $c$ depends on the shape of the distribution of $w$
and is defined as $c=<w^2>^2/(<w^4>-<w^2>^2)$. The scaling of the
signal to noise with the number of pixels is $(1-\beta)^4$ as
figure \ref{future} shows. It is the fact that we are
looking at the difference in the variance of $w$ for models
constructed to have the same $\bar w$, what makes distinguishing the
models so difficult as $\beta \rightarrow 1$. This fact 
is what forces $N_q$ to scale as $1/(\beta-1)^4$ instead of
$1/(\beta-1)^2$. 

Finally with more data we could not only get better constraints on
model parameters such as $\beta$ but also constrain more ambitious
models. For example, rather than have fixed size bubbles we could draw
the bubble sizes from some distribution or allow for correlations
between their positions. Moreover with more data one has to study in
detail certain sources of systematic error in our method like
contamination by metal lines or the effect of evolution along the
spectra.

\section{Conclusions}\label{conclusions}

The uniformity of the temperature in the IGM is an untested assumption
in most theoretical modelling of the Lyman $\alpha$ forest. We have
presented a statistical method to test this assumption. Our method
uses the small scale power of the forest flux as a thermometer and
searches for fluctuations in this power. As a result the method is
statistical in nature. An important advantage of the method we suggest
is that it does not require a direct comparison with numerical
simulations. It searches for changes in the statistical properties in
different parts of the spectra.

We illustrated our method with Q1422+231. We assumed
that in different regions of the spectrum the IGM could be in one of two
temperatures. We found that the ratio of small scale power in these
two type regions could not differ by more than a factor of $3.5$ if
the two type of regions occupy a comparable fraction of the spectrum.
This ratio of power translates in a factor of approximately $2.5$ in
temperature. If the transition between the two regimes is sharp, the
constraints are more stringent, the ratio of powers cannot differ by
more than a factor of $2$.

We have calculated how much data is needed to improve this constraints
substantially. We showed that ten quasars are needed to constrain
fluctuations of factors of two in the power, or equivalently factors
of $1.7$ in temperature. Furthermore with more data, more ambitious
models for the fluctuations can be considered.

{\bf Acknowledgements} The author acknowledges very useful comments
and suggestions from Joop Schaye and John Bahcall. Support for this
work is provided by the Hubble Fellowship HF-01116-01-98A from STScI,
operated by AURA, Inc. under NASA contract NAS5-26555.

\end{document}